\documentclass[acus]{JAC2003}
\usepackage{graphicx}
\usepackage{booktabs}
\begin{document}
\title{MgB$_{2}$ Nonlinear Properties Investigated Under Localized High RF Magnetic Field Excitation}
\author{Tamin Tai\thanks{tamin@umd.edu}, B. G. Ghamsari, Steven M. Anlage,
Center for Nanophysics and Advanced Materials,\\ Physics Department,
 University of Maryland, College Park, MD 20742, USA. \\ C. G.
Zhuang, X. X. Xi, Physics Department, Temple University,
Philadelphia, PA 19122 , USA}

\maketitle

\begin{abstract}
In order to increase the accelerating gradient of Superconducting
Radio Frequency (SRF) cavities, Magnesium Diboride (MgB$_{2}$) opens
up hope because of its high transition temperature and potential for low surface
resistance in the high RF field regime. However, due to the presence
of the small superconducting gap in the $\pi$ band, the nonlinear
response of MgB$_{2}$ is potentially quite large compared to a
single gap s-wave superconductor (SC) such as Nb. Understanding the
mechanisms of nonlinearity coming from the two-band structure of
MgB$_{2}$, as well as extrinsic sources, is an urgent requirement. A localized and strong RF
magnetic field, created by a magnetic write head, is integrated into
our nonlinear-Meissner-effect scanning microwave microscope \cite{T.
Tai}.  MgB$_{2}$ films with thickness 50 nm, fabricated by a hybrid
physical-chemical vapor deposition technique on dielectric
substrates, are measured at a fixed location and show a strongly
temperature-dependent third harmonic response. We propose that at
least two mechanisms are responsible for this nonlinear response,
one of which involves vortex nucleation and penetration into the
film.
\end{abstract}

\section{INTRODUCTION}
The discovery of superconductivity in MgB$_{2}$ in January 2001
\cite{J. Nagamatsu} ignited enthusiasm and interest in exploring its
material properties. Several remarkable features, for example a
high transition temperature ($T_{c}\sim $ 40 K ), a high critical
field, and a low RF surface resistance below $T_{c}$, shows great potential in
several applications such as superconducting wires and magnets. The
success of making high quality epitaxial MgB$_{2}$ thin films
provides another promising application as an alternative material
coating on superconducting radio frequency (SRF) cavities \cite{Xi}.
Over the past decade, the improvement of accelerating gradient in
Niobium (Nb) SRF cavities has almost reached the BCS limit, 57
$MeV/m$ \cite{TESLA}. In order to go further, new high $T_c$
materials with low RF resistance are required for interior coating
of bulk Nb cavities. High quality MgB$_{2}$ thin films may satisfy
the demands for SRF coating materials because these high quality
films can avoid the weak link nonlinearity between grains, and lead
to the possibility of making high-Q cavities \cite{Tajima}.

However, there still exist mechanisms that produce non-ideal behavior at low temperatures under high RF magnetic fields, such
as vortex nucleation and motion in the film \cite{A. Gurevich}.
In addition, due to the $\pi$ band and $\sigma$ band, the intrinsic
nonlinear Meissner effect of MgB$_{2}$ is large compared to other single-gap
s-wave superconductors \cite{G. Cifariello}. Therefore the study of
MgB$_{2}$ microwave nonlinear response in the high frequency region
(usually several GHz in SRF applications) can reveal the
dissipative and nondisipative nonlinear mechanisms and allow application of
these high quality MgB$_{2}$ films as cavity coatings. \\

In our experiment the localized harmonic response of superconductors is
excited by a magnetic write head probe extracted from a commercial
magnetic hard drive \cite{T. Tai}. Based on the gap geometry of the
magnetic write head probe, sub micron resolution is expected. We
present our observation of at least two measurable nonlinear
mechanisms involved in high quality MgB$_{2}$ films below $T_{c}$.
These films were grown on sapphire substrates by hybrid
physical-chemical vapor deposition technique (HPCVD). A detailed
description of the growth technique has been reported before \cite{X.
Zeng}. Finally, experimental nonlinearity data will be interpreted
as a combination of intrinsic nonlinear response~\cite{T. Dahm} and
vortex nonlinearity \cite{A. Gurevich}.

\begin{figure}[Experiment-technique]
   \centering
   \includegraphics*[width=2 in, angle=-90]{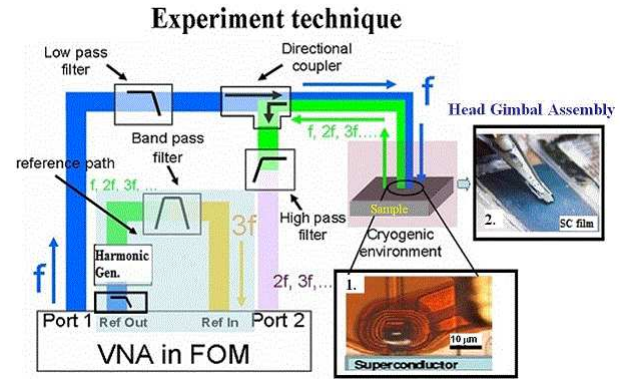}
   \caption{Set up of phase-sensitive measurement in nonlinear microwave microscopy. The frequency offset mode (FOM) of a vector network analyzer (VNA :model PNA-X N5242A) is used in this measurement.
   Inset 1. is a schematic picture of the probe above the sample and inset 2 shows the magnetic write head probe assembly on top of a superconducting thin film.}
   \label{figure_setup}
\end{figure}

\section{EXPERIMENTAL SETUP}
The experimental setup for amplitude and phase measurements of the
superconductor harmonic response is shown in Fig.
\ref{figure_setup}. An excited wave (fundamental signal) at
frequency $f$ comes from the vector network analyzer (VNA) and is
low-pass filtered to eliminate higher harmonics of the source
signal. This fundamental tone is sent to the magnetic write head
probe to generate a localized RF magnetic field on the
superconductor sample. Two insets in Fig. \ref{figure_setup} shows
close-up views of our magnetic write head probe on superconducting
samples.  Due to the intense nature of this field, the
superconductor responds by generating currents at both the
fundamental frequency and at harmonics of this frequency.  The
generated harmonic signal is high-pass filtered to remove the
fundamental signal $V_f$ and an un-ratioed measurement of $V_{3f}$
is performed on port 2 of the VNA. In order to get a phase-sensitive
measurement of the $3^{rd}$ harmonic signal coming from the
superconducting sample, a harmonic generation circuit is connected
to provide a reference $3^{rd}$ harmonic signal, and the relative
phase difference between the main circuit and reference circuit is
measured. Further detail about this phase-sensitive measurement
technique can be found in Ref. \cite{D. Mircea}. In this way we
measure the complex third harmonic voltage of $V_{3f}^{sample}(T)$
or the corresponding scalar power $P_{3f}^{sample}(T)$. The lowest noise
floor in our VNA is -145 dBm for the un-ratioed power measurement. A
ratioed measurement of the complex $V_{3f}^{sample}(T)/V_{3f}^{ref}$
is also performed at the same time. In this paper we only discuss
the unratioed measurements of $P_{3f}^{sample}(T)$ and qualitatively
discuss the mechanisms of third harmonic response of the MgB$_2$
film.

\section{THIRD ORDER NONLINEAR MEASUREMENT RESULTS}
\begin{figure}[tb]
    \centering
    \includegraphics*[width=2.7in]{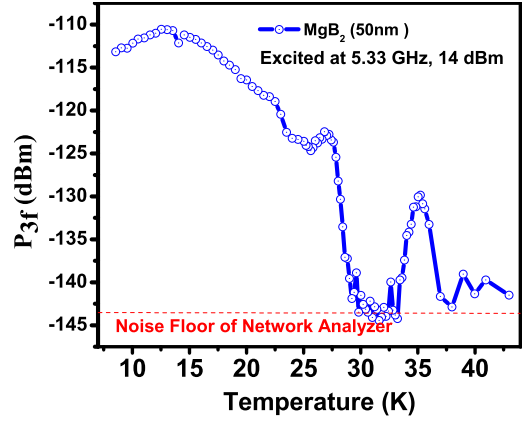}
    \caption{Temperature dependence of 3rd harmonic power $P_{3f}$ from a 50 nm thick MgB$_{2}$
    measured with an excited frequency of 5.33 GHz.}
    \label{MgB2_50nm}
\end{figure}

The measurement of the $3^{rd}$ order harmonic power ($P_{3f}$) is
performed near the center of the epitaxial MgB$_{2}$ film of
thickness 50 nm.  The $T_c$ of this sample is 36 K as measured by
the four point resistance method. Figure \ref{MgB2_50nm} shows the
temperature dependent $P_{3f}(T)$ curves at the excited frequency
5.33 GHz and excited power +14 dBm.  Above 40 K a very small signal
begins to arise above the noise floor of the network analyzer. This
$P_{3f}$ is from the magnetic write head probe itself. We have
measured the $P_{3f}$ of the magnetic probe on the surface of a bare
sapphire substrate and in general this probe nonlinearity is
negligible at excited powers under 14 dBm. Although excited powers
above +14 dBm excites stronger nonlinearity from the probe, this
nonlinearity is almost temperature independent in the Helium cooling
temperature range. Therefore probe nonlinearity can be treated as a
constant background signal above the noise floor of the spectrum
analyzer.  The mechanism of probe nonlinearity is the
hysteretic behavior of the yoke material \cite{Bean} and has been
discussed previously \cite{T. Tai}.

From Fig. \ref{MgB2_50nm}, a clear $P_{3f}(T)$ peak centered at 35 K shows up
above the noise floor.  This peak arises from the intrinsic nonlinear Meissner Effect (NLME) at $T_c$ due to the enhanced sensitivity of superconducting properties as the superfluid density decreases to near-zero levels. This peak at $T_{c}$ is also phenomenologically
predicted by Ginzburg-Landau theory, and is discussed further below.

We also note the onset of a low temperature nonlinearity below 27
K, which implies that another temperature dependent nonlinear
mechanism is active.  It may be that the applied RF field from the probe is strong enough to
penetrate into the superconductor and create deep flux penetration or even Abrikosov vortices
in a localized area. This new nonlinear mechanism dominates the
overall measured nonlinearity in the low temperature region. Further
qualitative discussion of the low temperature nonlinearity from a possible
Abrikosov vortex critical state will be addressed in detail below.

In addition, in the temperature regime of $29 K \sim 33 K$, there is
a minimum $P_{3f}$ signal, which implies no strong nonlinearity
mechanisms in this temperature range. The supercurrent of a vortex
circulates around the normal core with an approximate size of the
magnetic penetration depth $\lambda (T)$. Therefore once $\lambda
(T)$ is bigger than the film thickness (50 nm in this case) above a
certain temperature, vortex penetration due to parallel magnetic
field will be suppressed \cite{E. Guyon}. This regime would be very
suitable to fabricate a low nonlinearity superconducting response in
a multi-layer superconductor / insulator structure \cite{Gurevich2}.

\begin{figure}
\centering
    \includegraphics*[width=1.6in]{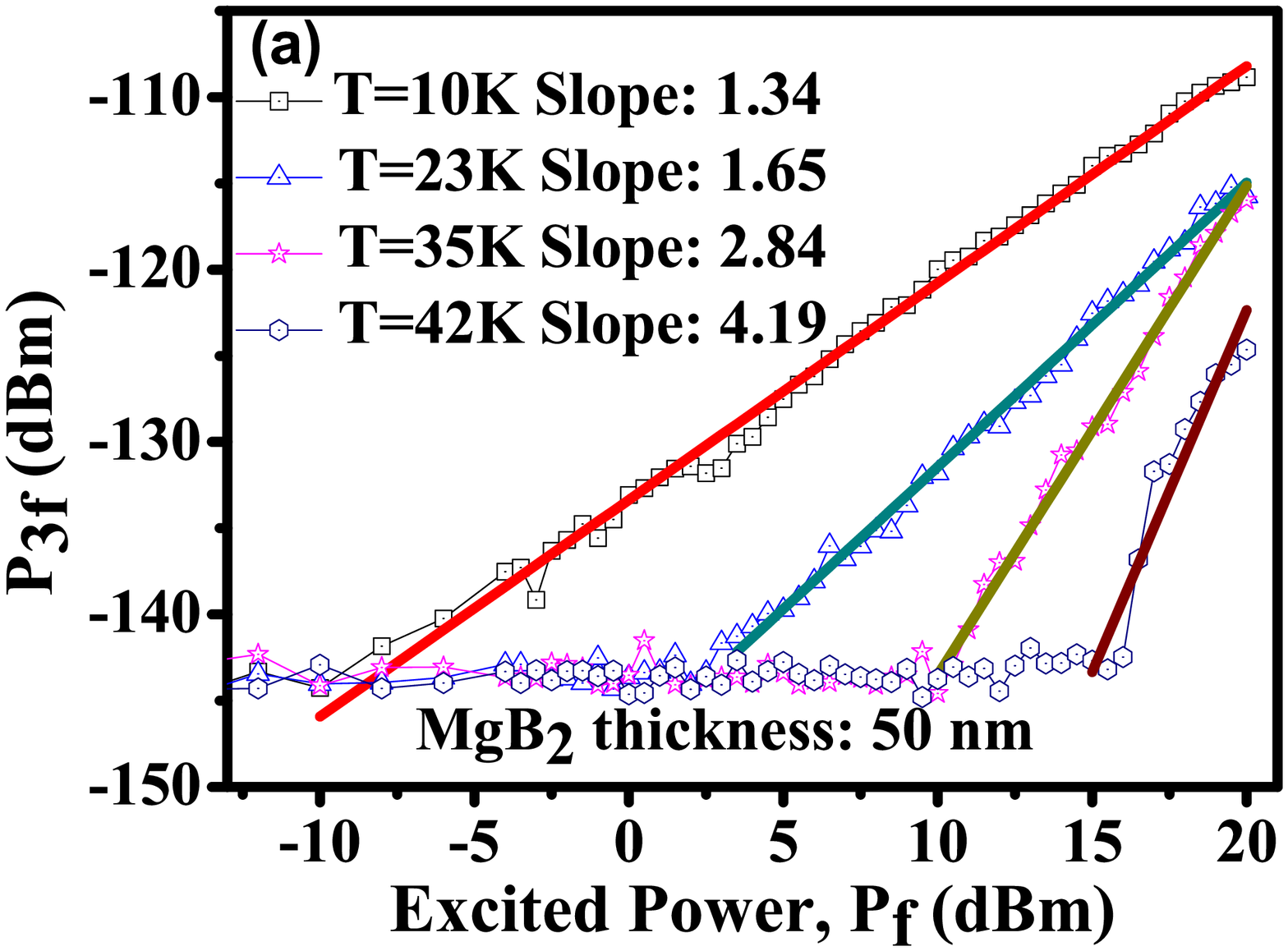}
    \includegraphics*[width=1.6in]{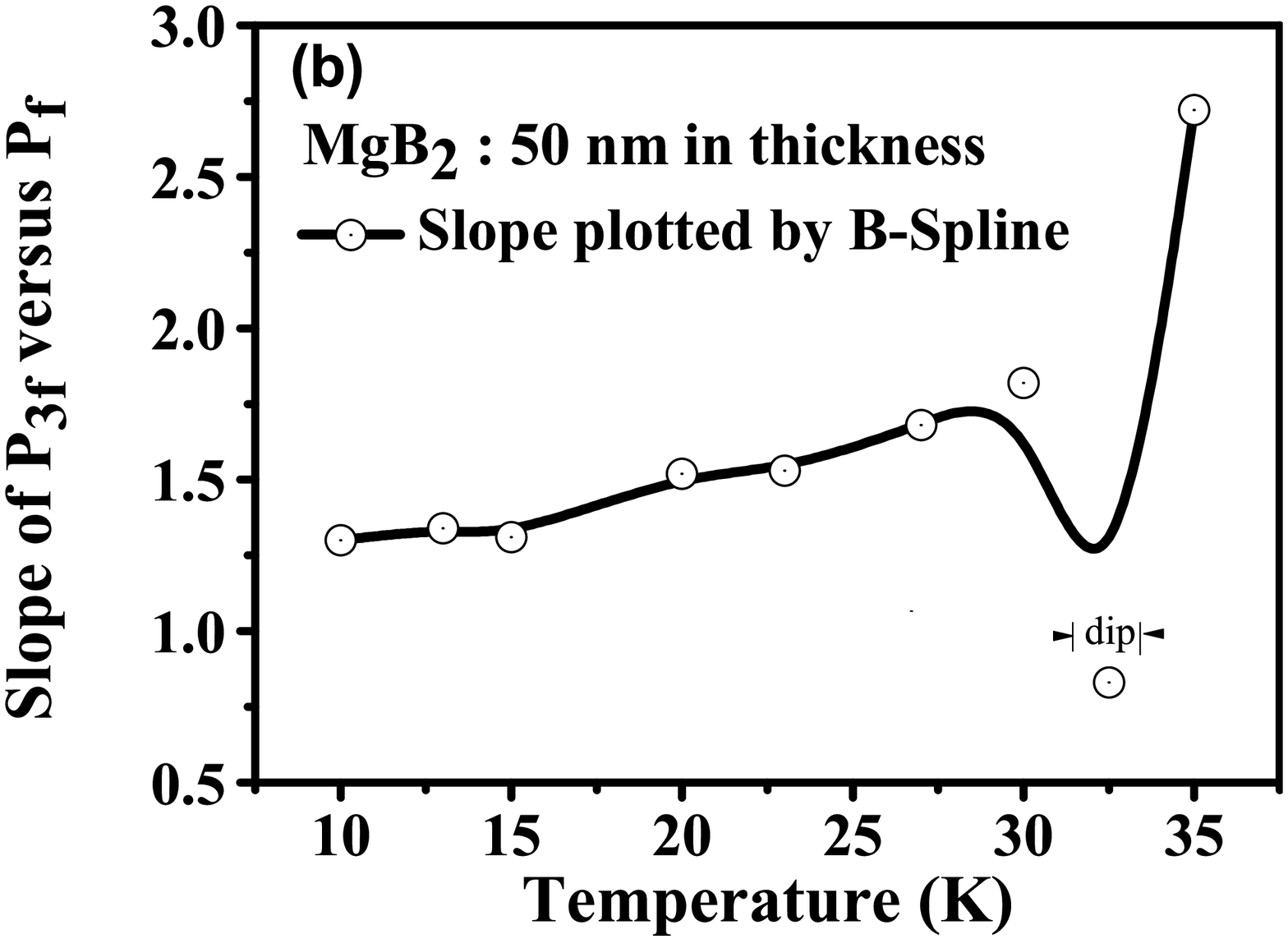}
    \caption{(a) Power dependence of P$_{3f}$ on P$_{f}$ for the 50 nm
    thick MgB$_{2}$ film. (b) Fitted slope at selected temperatures for the film.
    The marked dip may describe a vortex-free region and its small slope is likely due to the probe nonlinearity.}
    \label{MgB2SlopeAll}
\end{figure}

Measurements of the dependence of $P_{3f}$ on $P_{f}$ are shown in
Fig. \ref{MgB2SlopeAll} (a) for the 50 nm thick MgB$_{2}$ film at
some selected temperatures. In the normal state of MgB$_{2}$, the
measured nonlinearity comes from the probe itself and shows a slope
steeper than 3 at high excited power above +15 dBm. In the intrinsic
nonlinear Meissner regime, the slope is 2.84, very close to 3 as
predicted for the intrinsic NLME \cite{John Lee}. Based in part on
this evidence, we believe that in the high temperature region close
to T$_{c}$, the $P_{3f}$ comes from the intrinsic NLME. In the low
temperature regime, the slopes of P$_{3f}$ vs. P$_{f}$ are around
1.5. This value is similar to that predicted by many
phenomenological models (between $1 \sim 2$) in an Abrikosov vortex
critical state \cite{D. E. Oates} \cite{J. Mateu}. It should be
noted that the low temperature nonlinearity can be easily excited at
low power. Figure \ref{MgB2SlopeAll} (b) show the P$_{3f}$-P$_{f}$
slope evolution from a flux/vortex dominated nonlinear regime at low
temperature to a NLME regime around $T_c$ for this MgB$_{2}$ film.

\section{Intrinsic NLME of MgB$_2$}
\begin{figure}[tb]
    \includegraphics*[width=3in]{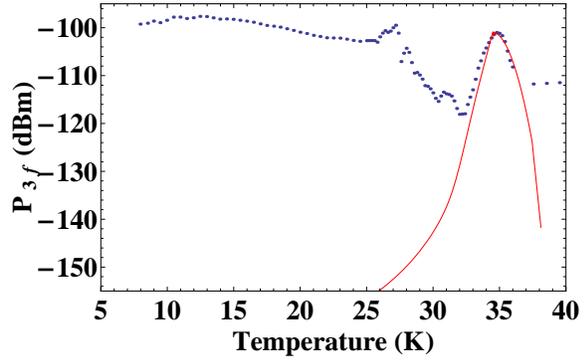}
    \caption{Solid red line is the simulated result of the intrinsic NLME of MgB$_{2}$ with thickness 50 nm
    under the assumption that the magnetic probe provides a field described by a geometry factor
    $\Gamma= 8.3*10^{5}$ $A^{3}/m^{2}$. Other parameters used in this calculation include $\lambda(0K)$=100
    $nm$, $\lambda_{cutoff}$=800 $nm$, $J_{cutoff}=4.2*10^{11}$ $A/m^2$ and $T_c$=34.6 $K $
    with a standard deviation of Gaussian spread of $\delta T_c$=1.3 $K$. The blue dots are the data from the experiment. }
    \label{MgB2NLME}
\end{figure}

The intrinsic nonlinearity comes from the backflow of excited
quasiparticles in a current-carrying superconductor, which results in an effective decrease of the superfluid density. Therefore, a two band quasiparticle
backflow calculation should be applied to the MgB$_{2}$ intrinsic
nonlinearity. Based on the work of Dahm and Scalapino\cite{T. Dahm}, the temperature
and induced current density dependent superfluid density
$n_{s}(T,J)$ can be written as

\begin{equation}\label{eq:superfluid}
    \frac{n_{s}(T,J)}{n_{s}(T,0)}=1-({J \over J_{NL}})^2 ;
    J_{NL}={J_{c,\pi} \over \sqrt{b_{\pi}(T)+b_{\sigma}(T){J_{c,\pi}^{2}
\over J_{c,\sigma}^{2}} }}
\end{equation}
where b$_{\sigma}$ and b$_{\pi}$ are the temperature dependent
nonlinear coefficients for the $\sigma$ band and $\pi$ band,
respectively, and their values are defined in reference \cite{T.
Dahm}. $J_{c,\sigma}=4.87\times10^8 A/cm^2$ and
$J_{c,\pi}=3.32\times10^8 A/cm^2$ are the pair-breaking current
densities for the two bands. For a 50 nm thick MgB$_2$ thin film,
the generated third harmonic power $P_{3f}(T)$ is estimated by
substituting $J_{NL}$ into the following equation \cite{John Lee}

\begin{equation}\label{eq:P3f}
    P_{3f}(T)=\frac{\omega^2\mu_{0}^2\lambda^4(T)\Gamma^2}{32Z_
    {0}d^{6}J^4_{NL}(T)}
\end{equation}
where $\omega$ is the angular frequency of the incident wave, $d$ is
the film thickness, $\lambda (T)$ is the temperature dependent
magnetic penetration depth, $Z_0$ is the characteristic impedance of
the transmission line in the microscope, and $\Gamma$ is geometry
factor which is estimated to be $10^5 A^3/m^2$ for the magnetic
write head field distribution under a 100 $mW$ excited power. The
solid red line in Fig. \ref{MgB2NLME} shows the $P_{3f}(T)$
simulated results of Eqs. (\ref{eq:superfluid}) and (\ref{eq:P3f})
for the 50 nm thick film at a 5.33 GHz excited frequency. This
intrinsic NLME response has measurable values above the noise floor
only in the high temperature region near $T{_c}$. The experimental
data of the MgB$_{2}$ film under a +18 dBm, 5.33 GHz microwave
excitation is shown in the blue dots. Therefore, at lower
temperatures the nonlinear mechanism must be of a different nature.

\section{Nonlinearity in Abrikosov Vortex Critical State}
\subsection{Nonlinearity From Moving Vortices}
Vortex nucleation and penetration into the film induces a dynamic
instability and generates harmonic response. The equation of motion
of a vortex in a semi-infinite superconductor driven by a harmonic magnetic field is given by \cite{A. Gurevich}
\begin{equation}\label{eq:vortex}
\eta\dot{x}=\frac{\Phi_{0}B_{0}}{\mu_0\lambda(T)}e^{i{\omega}t}e^{-x/\lambda(T)}-\frac{\Phi_{0}^{2}}
{2\pi\mu_{0}\lambda^{3}(T)}K_{1}(\frac{2\pi x}{\lambda(T)})
\end{equation}
where $x$ is the coordinate of the vortex position with respect to
the surface ($x$=0), $\eta$ is the Bardeen-Stephen vortex viscosity,
$\Phi_0$ is the flux quantum, $\omega$ is the angular frequency of
the incident wave, $\mu_0$ is the permeability of vacuum, $B_0$ is
the magnitude of RF magnetic field on the SC surface and $K_1(x)$ is
the modified Bessel function. The first term on the right hand side
is the Lorentz force per unit length on the vortex due to the
screening currents created by the driving field. The second term on
the right is the force per unit length exerted by the image vortex
that arises from the SC/vacuum surface. This equation assumes a bulk
superconductor.

The solution for the trajectory of this single vortex
is shown in Fig. \ref{vortex_dynamics} as a function of
time in the lower solid blue curve. The applied RF field $B(t)$ is also
included in the figure to illustrate the relation of the vortex
position and the applied field with time. The time for the first
vortex entry can be determined as \cite{A. Gurevich},
\begin{equation}\label{nucleationT}
t_0=\frac{\arcsin(\frac{B_v}{B_0})}{\omega}
\end{equation}
where $B_v$ is the penetration field of a vortex (assuming $B_v$ $<$
$B_0$). A vortex will start to nucleate and enter into the film when
$B(t)$ exceeds the Bean-Livingston barrier \cite{Bean-Livingston}.
This vortex also creates a supercurrent circulating around the core
and distorts the Meissner screening current near the surface. During
the reverse part of the RF cycle, the Meissner screening current is
enhanced so that at time $t_c$ an anti-vortex will penetrate into
the superconductor as shown by the dashed red line. This second vortex will
annihilate with the first vortex at time $t_a$. This procedure of
vortex-antivortex entry and annihilation continues and will generate
a third harmonic signal.

For further quantitative modeling,
the following two cases should been
taken into consideration:

\begin{figure}
    \centering
    \includegraphics*[width=2in, angle=-90]{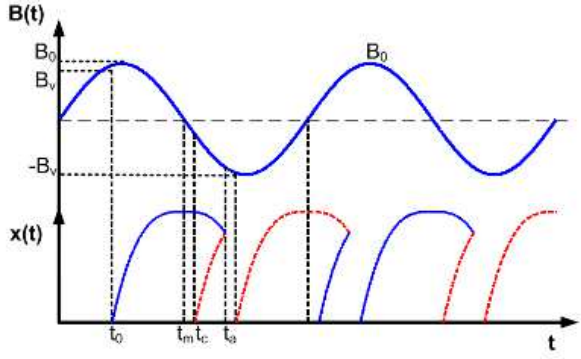}
    \caption{A calculation of vortex position $x(t)$ with the applied RF field $B(t)$ over several RF cycles.
    Here $t_{0}$ indicates the time of vortex entry, $t_{m}$ is the time when the vortex reaches its maximum distance into the film, $t_{c}$ is the antivortex nucleation time, and $t_{a}$ is the vortex / anti-vortex collision time.}
    \label{vortex_dynamics}
\end{figure}

(1) Finite thickness of the film. \\
The vortex equation of motion given above is only
suitable for bulk materials and assumes that a uniform magnetic field is
applied parallel to the SC surface. In the finite thickness case, an
infinite number of image vortices are required to satisfy the
boundary conditions. However we can make an approximation that only
two image vortices are required. Therefore the equation of motion of
the vortex is modified to,
\begin{eqnarray}\label{eq:vortex2}
\eta\dot{x}=\frac{\Phi_{0}B_0}{\mu_0\lambda}e^{i{\omega}t}e^{-x/\lambda}-\frac{\Phi_{0}^{2}}{2\pi\mu_{0}\lambda^{3}}[K_{1}(\frac{2\pi
x}{\lambda}) \nonumber \\
-K_{1}(\frac{2\pi(d-x)}{\lambda})]
\end{eqnarray}
where $d$ is the thickness of the film.
This modification for a second image force will help improve the quantitative modeling.\\

(2) surface roughness of the film \\
In the Bean-Livingston model the superconducitng surface is assumed to be a perfect plane~\cite{Bean-Livingston}. When the surface has roughness with characteristic length $\geq$ $\xi$ (coherence length), a
geometry effect should be taken into consideration \cite{Ernst
Helmut Brandt}. Generally, for a sharper corner, the Meissner
screening current density will be enhanced and the penetration field of
the first vortex entry ($B_{v}$) will decrease. For example, at a
corner with a $90^0$ angle, an enhancement of the screening current is
roughly estimated to be a factor of 4 \cite{Ernst Helmut Brandt}. This
means a vortex will penetrate at sharp points or cusps easily and
reduce the vortex nucleation time during the RF cycle. Therefore, nonlinear harmonic response
will be increased compared to the case of a perfect plane. Hence for a given excitation level, the harmonic response will depend on the surface topography, and an image showing this contrast can be built up by raster scanning the magnetic probe.\\

\subsection{Nonlinearity From Switching Between the Meissner State and the Vortex Critical State}
In addition to vortex and antivortex nucleation and motion, another possibility to
generate a $P_{3f}$ in the Abrikosov vortex critical state is the
switching between this state and the nonlinear Meissner state. While the peak value of the applied RF
magnetic field is higher than the surface penetration field of the
superconductor, the material will switch into the critical vortex
state from the Meissner state.  This process of switching
between states implies
another source of nonlinear harmonic response.

\begin{figure}
    \centering
    \includegraphics*[width=1.5in, angle=-90]{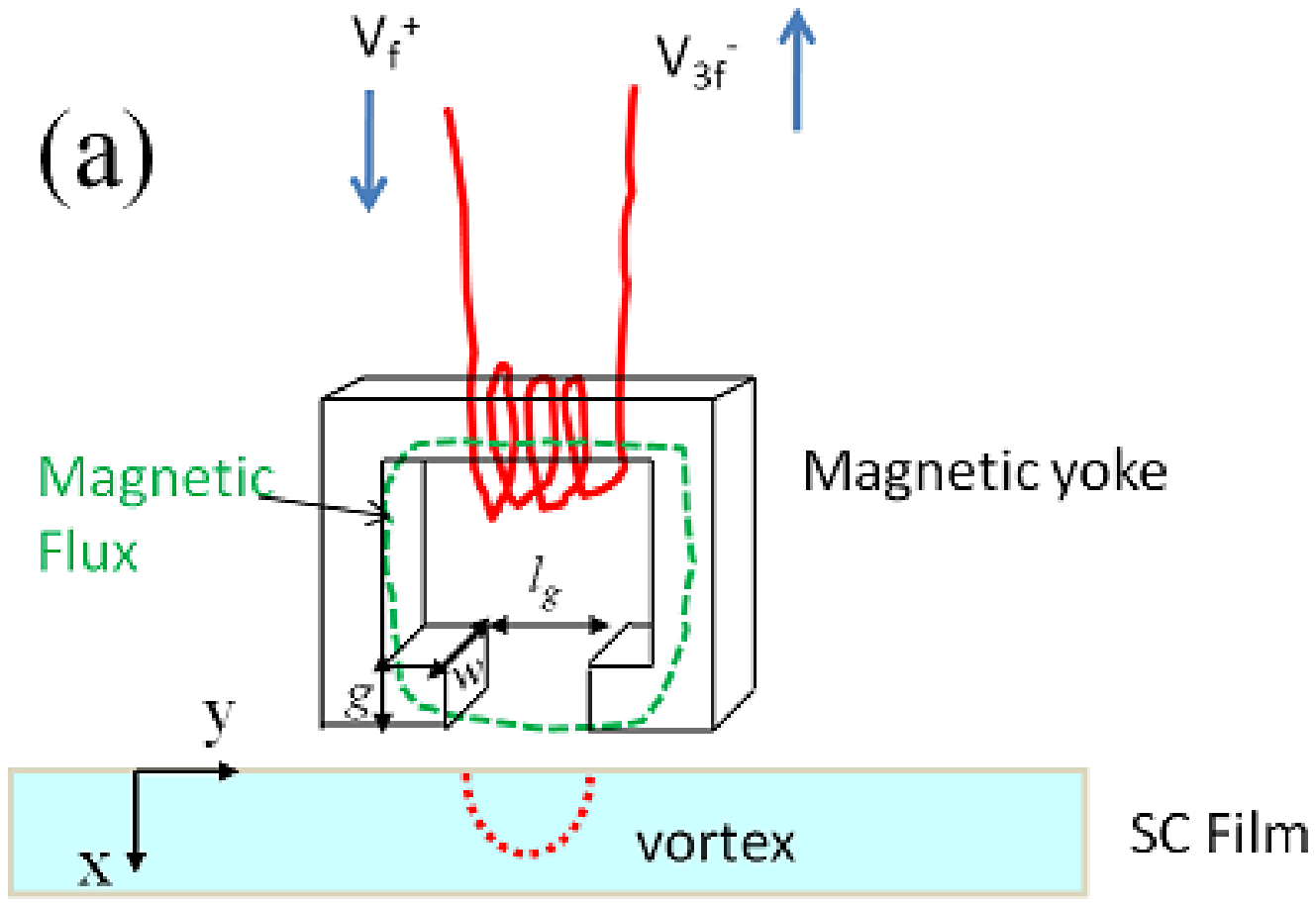}
    \includegraphics*[width=1.5in, angle=90]{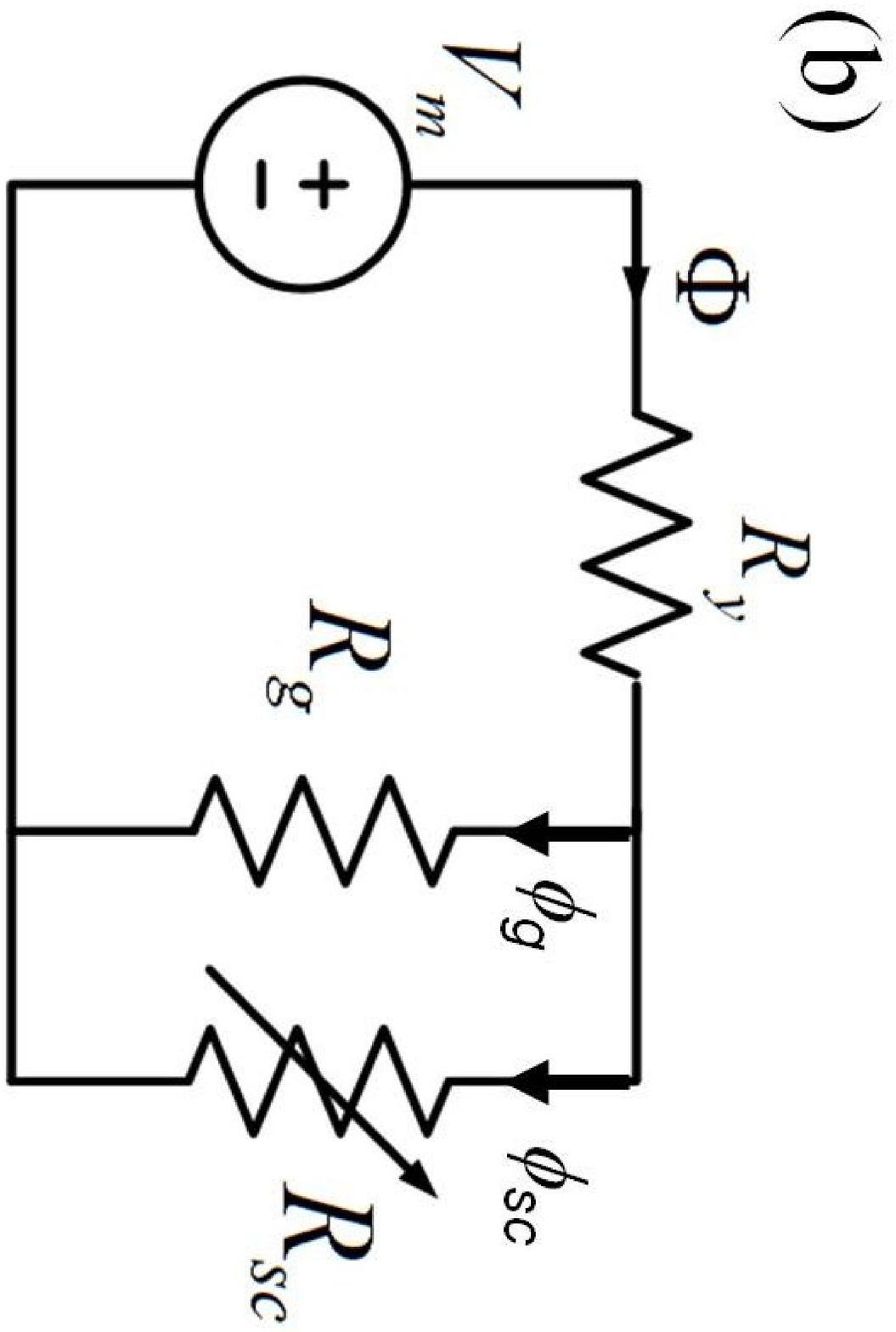}
    \caption{(a) Schematic illustration of the magnetic flux coming from the yoke to the
    superconductor where $w,g$ and $l_g$ represent the width, the thickness, and
    the length of the gap, respectively.  The length $l_g$ is on the order of 200 nm for our write head probe.
    (b) An equivalent magnetic circuit for the magnetic flux transport from the yoke to the superconductor.
    Note $R_{sc}$=$R_s$+$R_v$ is a time variable reluctance..}
    \label{Mcircuit}
\end{figure}

Fig. \ref{Mcircuit} (a) shows a schematic illustration of our
experiment in which the RF magnetic field from the magnetic write
head probe interacts with the superconductor underneath the probe.
One can model the flux distribution with an equivalent magnetic
circuit as shown in Fig.\ref{Mcircuit} (b). The inductively coupled
driving line provides a magnetomotive force $(V_m)$ to the yoke with
a reluctance $R_{y}$. A magnetic flux $\Phi$ is channeled down along
the yoke to the gap.  There the flux can divide into two branches:
one directly goes through the gap with a reluctance $R_g$ and the
other shunts into the superconductor with a reluctance $R_{sc}$. The
reluctance of superconductor $R_{sc}$ is a time-variable reluctance. It is a combination of the reluctance $R_s$ from the nonlinear Meissner
state and the reluctance $R_v$ from the vortex critical state. While
the applied field $B(t)$ is smaller than the penetration field
$B_v$, the reluctance will remain at the value of $R_s$. Once $B(t)$
$\geq$ $B_v$, an additional reluctance channel $R_v$ is created.  Whether the vortex enters as a semi-loop (as assumed above), or as a vortex-antivortex pair, remains to be evaluated.

Because a magnetic
circuit is analogous to an electric circuit, we can compute $R_v$ and
$R_s$ with node-voltage analysis. Assume that the flux going
through $R_{g}$ and $R_{sc}$ is $\phi_{g}$ and $\phi_{sc}$,
respectively. In the nonlinear Meissner state, we obtain
\begin{equation}\label{circuit}
\phi_g=B_0 A_{gap}=B_0(w*g)  \hspace{3mm};\hspace{3mm} \phi_{sc}=B_0
(w*\lambda)
\end{equation}
where $A_{gap}$ is the cross-sectional area of the gap, $w$ and $g$ are
the width and the thickness of the gap, respectively (see Fig. \ref{Mcircuit} (a)), and $\lambda$ is
temperature dependent penetration depth. Applying the node-voltage
law, we have $\phi_g R_g=\phi_{sc} R_s$. Finally, the reluctance of the
nonlinear Meissner state is given by
\begin{equation}\label {R_s}
R_s=\frac{g}{\lambda} R_g  \qquad R_g=\frac{l_g}{\mu_0 A_{gap}}
\end{equation}
where $l_g$ is the length of the gap (see Fig. \ref{Mcircuit} (a))
and $\mu_0$ is the permeability of vacuum. On the other hand, in
the vortex critical state, we have the magnetomotive force as
\begin{equation}
\phi_0 R_v=\int_{gap}{ H_0 dy}=\frac{B_0 l_g}{\mu_0} \qquad
\phi_0=B_0 \lambda^2
\end{equation}
where $\phi_0$ is the flux quantum and the latter equation assumes that the vortex carries one unit of magnetic flux, which may not always be the case. Hence the reluctance of the vortex
critical state is $R_v=\frac{l_g}{\mu_0 \lambda^2}$. In addition to the nonlinearity of the penetration depth with RF field, we believe the
transient between $R_v$ and $R_s$ will also induce a third harmonic response.\\

\section{CONCLUSIONS}
A strongly temperature-dependent third harmonic response is found in
high quality MgB$_2$ films. In addition to the intrinsic
nonlinearity, the nonlinearity coming from the Abrikosov critical
state may also be involved. From the dependence of P$_{3f}$ on
P$_{f}$, the nonlinearity mechanism changes from a intrinsic
nonlinear Meissner effect to a possible vortex critical state dominated
nonlinearity upon cooling the high quality epitaxial MgB$_2$ film.
The mechanics of nonlinearity in the Abrikosov vortex critical state
can be qualitatively interpreted by two models - first:
annihilation of moving vortex $\&$ antivortex pairs and second: state
switching between a Meissner state and a vortex critical
state.

\section{Acknowledgement}
This work is supported by the US Department of Energy/ High Energy
Physics through grant $\#$ DESC0004950, and also by the ONR AppEl,
Task D10, (Award No.\ N000140911190), and CNAM. The work at Temple University is supported by DOE under grant No. DE-SC0004410.

\end{document}